\documentclass{article}
 \input{epsf}
\usepackage{amsmath}
\usepackage{epsfig}
\usepackage{amssymb,amsfonts}

\begin{document}

\title{The non-compact elliptic genus: mock or modular}

 \author{
 Jan Troost
}
 
\maketitle

\begin{center}
 \emph{Laboratoire de Physique Th\'eorique}\footnote{Unit\'e Mixte du CNRS et
    de l'Ecole Normale Sup\'erieure associ\'ee \`a l'universit\'e Pierre et
    Marie Curie 6, UMR
    8549. LPTENS-10/16} \\
\emph{ Ecole Normale Sup\'erieure  \\
24 rue Lhomond \\ F--75231 Paris Cedex 05 \\ France}
\end{center}

\begin{abstract}
  We analyze various perspectives on the elliptic genus of non-compact
  supersymmetric coset conformal field theories with central charge
  larger than three.  We calculate the holomorphic part of the
  elliptic genus via a free field description of the model, and show
  that it agrees with algebraic expectations. The holomorphic part of
  the elliptic genus is directly related to an Appell-Lerch sum and
  behaves anomalously under modular transformations. We
  analyze the origin of the anomaly by calculating the elliptic genus
  through a path integral in a coset conformal field theory. The path
  integral codes both the holomorphic part of the elliptic genus, and
  a non-holomorphic remainder that finds its origin in the continuous
  spectrum of the non-compact model. The remainder term can be shown
  to agree with a function that mathematicians introduced to
  parameterize the difference between mock theta functions and Jacobi
  forms.  The holomorphic part of the elliptic genus thus has a path
  integral completion which renders it non-holomorphic and
  modular. \end{abstract}

\section{Introduction}
The elliptic genus
\cite{Schellekens:1986yi,Schellekens:1986yj,Schellekens:1986xh,Witten:1986bf}
is a generalized index that codes information on the spectrum of $N=2$
superconformal field theories in two dimensions. It has applications
in the calculation of anomalies  and threshold
corrections in string theory, in algebraic
geometry, in the theory of modular forms and in the microscopic
calculation of black hole entropies.

A basic example of an elliptic genus is the elliptic genus of the
compact $N=2$ minimal superconformal field theories.  These conformal
field theories have a representation as the infrared fixed point of
supersymmetric Landau-Ginzburg models
\cite{Martinec:1988zu,Vafa:1988uu}, or alternatively as a gauged
Wess-Zumino-Witten model \cite{Di Vecchia:1985iy,Di Vecchia:1986kw,Di
  Vecchia:1986ts}.  For these prototypical compact superconformal
field theories, it is possible to calculate the elliptic genus in at
least three ways.  The elliptic genus of other compact models can then
be computed from that basic building block for instance by taking
orbifolded products. One can often compare the algebraic results to
geometric calculations of elliptic genera.

The three ways to compute the elliptic genus of the $N=2$ minimal
models are as follows. The first is by identifying the spectrum of the
model \cite{Di Vecchia:1985iy,Di Vecchia:1986kw,Di
  Vecchia:1986ts,Boucher:1986bh,Dobrev:1986hq}, and in particular the
representations that appear in a modular invariant partition function
\cite{Cappelli:1987xt,Gepner:1986hr}. The elliptic genus is then the
spectrum of left-movers (coded in a sum of irreducible characters)
with the right-movers in a Ramond ground state.  A second way to
compute the elliptic genus is by a free field calculation
\cite{Witten:1993jg} based on the fact that the $N=2$ minimal models
are the infrared fixed point of supersymmetric Landau-Ginzburg models
\cite{Martinec:1988zu,Vafa:1988uu}. The calculation gives rise to an
alternative expression for the elliptic genus which was shown to agree
with the algebraic formula \cite{DiFrancesco:1993dg}.  A third way of
computing the elliptic genus is via the description of the model as a
gauged Wess-Zumino-Witten model \cite{Henningson:1993nr}. The fields
of the model can be shown to acquire the same transformation
properties as the free fields of the Landau-Ginzburg model, and
moreover the path integral localizes, leading to an identical free
field calculation as the one performed in the Landau-Ginzburg
description.  There are many applications to more complicated models
based on basic building blocks of central charge smaller than three.

Our aim in this paper is to get a firmer grip on the elliptic 
genus\footnote{We will use the term elliptic genus for the twisted partition
function of the theory as defined by a path integral twisted by
$U(1)$ R-charge. An alternative would be to reserve this term
for the purely holomorphic part of the twisted partition function.} of
a non-minimal $N=2$ superconformal field theory which lies outside the
above class of theories. We will deepen our understanding of the
elliptic genus of building blocks with central charge larger than
three from the three perspectives discussed above.

For now, the proposed elliptic genus
is based on the calculation of the partition function of the bosonic
$SL(2,\mathbb{R})$ coset model \cite{Hanany:2002ev}. The
supersymmetric generalization was performed in
\cite{Eguchi:2004yi,Israel:2004ir}.  This allows for a proposal for
the elliptic genus as the algebraic sum of discrete characters in the
Ramond sector \cite{Eguchi:2004yi,Eguchi:2004ik}. Only discrete
characters are assumed to contribute, since only those allow for a
Ramond sector ground state (for the right-movers). However, the
regularization used in both \cite{Hanany:2002ev} and
\cite{Eguchi:2004yi,Israel:2004ir} is not modular invariant.  In
\cite{Israel:2004ir} is was also shown that keeping track of the
multiplicities of descendent states necessitates a more covariant
treatment. An unexploited idea is to regulate the volume divergence of
the non-compact model covariantly by subtraction of the asymptotic
linear dilaton spectrum. In the absence of such an analysis, the fact
that the proposed elliptic genus transforms non-covariantly under
modular transformation properties is not understood. Though it is
known from the mathematics literature how to patch up this annoying
feature of the holomorphic part of the elliptic genus
\cite{Zwegers,Eguchi:2008gc}, the physical origin of the patchwork
is obscure.

In this paper, we first strengthen the motivation for the holomorphic
part of the elliptic genus through a free field
calculation. Secondly, we derive the holomorphic part
of the elliptic genus from a path integral calculation, and identify its
remainder, thus gaining insight into its modular properties. Indeed, we will
identify a non-holomorphic part of the elliptic genus that is
necessary to complete the elliptic genus into a Jacobi form. 

This is a neat addition to our understanding of the modular properties
of non-rational conformal field theories (see
e.g. \cite{Miki:1989ri,Eguchi:2003ik,Israel:2004xj}), as well as the
isolation of discrete states from the continuum in a chiral sector of
conformal field theory \cite{Benichou:2008gb}.  Our analysis is also
related to the modular properties of the characters of superconformal
algebras \cite{Semikhatov:2003uc}, invariants of three-manifolds, and
number theory \cite{Zagier}, the entropy of Calabi-Yau manifolds
\cite{Eguchi:2010xk}, and to the entropy of black holes and the
crossing of walls of marginal stability
\cite{Gaiotto:2008cd}\cite{Manschot:2009ia}\cite{DMZ}.

\section{Two perspectives on the holomorphic part of the elliptic
  genus}
\subsection{The free field perspective}
Recall that the elliptic genus is the trace over the Hilbert space of
an $N=2$ superconformal field theory, weighted by the
fermion number, the left $U(1)_R$ charge and the conformal dimensions
of the states:
\begin{eqnarray}
 \mbox{Tr} (-1)^F z^{J_0^R} q^{L_0-\frac{c}{24}} \bar{q}^{\bar{L}_0-\frac{c}{24}}.
\end{eqnarray}
Under suitable
conditions on the Hilbert space, the trace projects onto right-moving
ground states.

In this section we give a free field derivation of (the holomorphic
part of) the elliptic genus of a superconformal field theory with
central charge $c$ greater than three and of the form $c= 3 +
\frac{6}{k}$ where $k$ is an integer.  The analysis is similar to the
compact case \cite{Witten:1993jg}, so we mainly concentrate on the differences.  We specify
the superconformal field theory in terms of a supersymmetric action
with a Liouville potential term. {From} the potential term, we derive
the necessary properties of the free field that allow us to then
compute the elliptic genus while neglecting the interactions due to the
potential.

We consider a two-dimensional $N=(2,2)$ supersymmetric quantum field
theory containing a free chiral superfield $\Phi$ with standard
kinetic term, supplemented with a superpotential F-term:
\begin{eqnarray}
S_{pot} &=& \mu \int d^2  \sigma \, d^2 \theta  \,  e^{\sqrt{\frac{k}{2}} \Phi}.
\end{eqnarray}
There is also a supersymmetrized coupling to the background worldsheet
Ricci scalar curvature which corresponds to a linear dilaton in the
direction $\Phi$ with slope $Q=\sqrt{2/k}$. It renders the
superpotential term $S_{pot}$ marginal.  The bosonic superfield $\Phi$
contains a complex dynamical bosonic field $\phi=-\rho+i\theta$, a
complex auxiliary field $F$ and two Weyl fermions $\psi_{\pm}$. We choose
the
field $\theta$ to have radius $R =
\sqrt{{2k}}$.\footnote{We work in units where the self-dual radius
is $\sqrt{2}$. Note that any integer multiple of
  the radius $R_{min}=\sqrt{2/k}$ 
is a consistent choice. In particular, for
our purposes here the
  radius $R_{min}=\sqrt{2/k}$ would be an equivalent choice,
 up to the interchange of winding and momentum in the following
discussion.}
For an $N=(2,2)$ supersymmetric model in two dimensions with a chiral
superfield $\Phi$, the supersymmetry transformation rules
read\footnote{We use the conventions and notations of
  \cite{Witten:1993yc} where the transformation rules of
  \cite{Wess:1992cp} were reduced to two dimensions.}:
\begin{eqnarray}
\delta \phi &=& \sqrt{2} ( \epsilon_+ \psi_- - \epsilon_- \psi_+) 
\nonumber \\
\delta \psi_+ &=& i \sqrt{2} (\partial_0 + \partial_1) \phi \bar{\epsilon}_-
- \sqrt{2} \epsilon_+   \sqrt{\frac{k}{2}} e^{ \sqrt{\frac{k}{2}} \phi^\ast}
\nonumber \\
\delta \psi_- &=& -i \sqrt{2} (\partial_0-\partial_1) \phi \bar{\epsilon}_+ - \sqrt{2} \epsilon_-  
 \sqrt{\frac{k}{2}}  e^{ \sqrt{\frac{k}{2}} \phi^\ast}.
\end{eqnarray}
The four supersymmetry parameters are
$\epsilon_{\pm},\bar{\epsilon}_{\pm}$. The derivatives with respect to
the coordinates $\sigma^{0,1}$ are denoted $\partial_{0,1}$. To
compute the elliptic genus, we must determine the charges of the
fields under the left-moving $U(1)_R$ group.  We assign charge $0$ to
the right-moving supersymmetry transformation parameter $\epsilon_-$
and charge $+1$ to the supersymmetry parameter $\epsilon_+$.  {From}
the supersymmetry transformation rules and the fact that:
\begin{eqnarray}
\delta e^{ \sqrt{\frac{k}{2}} \phi} 
& =&  \sqrt{{k}} e^{ \sqrt{\frac{k}{2}} \phi}  ( \epsilon_+ \psi_- - \epsilon_- \psi_+),
\end{eqnarray}
we conclude that the left-moving $U(1)_R$ acts on the fields as follows:
\begin{eqnarray}
\psi_- & \rightarrow & e^{- i \beta} \psi_-
\nonumber \\
e^{ \sqrt{\frac{k}{2}} \phi} & \rightarrow & e^{+ i \beta} e^{ \sqrt{\frac{k}{2}} \phi} 
\nonumber \\
\psi_+  & \rightarrow & \psi_+.
\end{eqnarray}
We draw our first conclusions:
\begin{itemize}
\item At radius $R=\sqrt{2k}$ we have left-moving momenta
  $p_{\theta,L} = \sqrt{\frac{1}{2k}} (n-kw)$ (and right-moving
  momenta $p_{\theta,R}= \sqrt{\frac{1}{2k}} (n+kw)$) where $n$ and
  $w$ are integer.
 \item The left-moving fermion has charge $-1$.  The right-moving fermion has zero charge. 
 \item The field $e^{i \sqrt{\frac{1}{2k}}\theta}$ (which for $k$
integer has the minimal left-moving momentum) has charge
   $+\frac{1}{k}$. The charge is carried by the zero mode of the
   angular component of the complex boson.
\end{itemize}

\subsection*{Remark}
To remain formally closer to the compact case, we could have written the superpotential term as
\begin{eqnarray}
W & = & Z^{-k}
\end{eqnarray}
where $Z = e^{-\frac{1}{\sqrt{2k}} \Phi}$. A crucial difference with
the compact case is that there is a singular region for the potential
near $Z=0$. Moreover, the potential leaves the field $Z$ free to
fluctuate at large values in field space. The first difference prompts
us to choose the exponential variable which is better behaved near the
origin, along with a standard kinetic term. This has the crucial
consequence that the configuration space is punctured, and that we can
have non-trivial winding configurations around the origin of
configuration space.  Moreover, it is now only the zero-mode of the
complex boson that carries charge.  The second difference forces us to
specify a linear dilaton behaviour at infinity (see
e.g. \cite{Hori:2001ax}), in order to obtain the right central charge
in the non-compact, free region of configuration space. Finally,  we note
that the exponential form of the potential
makes it manifest that one
cannot
continuously dial the coupling constant to zero.

\subsection*{Free field proposal}
We compute the free field elliptic genus in two steps. First, let us
consider the subsector of the Hilbert space in which the left- and
right-moving momentum of the compact boson $\theta$ are equal.  In
other words, we are in the zero winding sector. In the sector of the
Hilbert space where we only allow for real Liouville momenta $p_\rho$,
the elliptic genus will be zero. When we allow for imaginary momenta
as well, then we can compensate the right-moving conformal dimension
of the operators $e^{i n \sqrt{\frac{1}{2k}}
  \theta}$ by the Liouville momentum
contribution, and obtain a right-moving ground state. In the
left-moving sector, due to the diagonal spectrum for the Liouville
mode, as well as the diagonal subsector we are in, these modes then
act as zero-modes with $U(1)_R$ charge $+n/k$. To compute the
partition function, we must regulate the contribution from these
left-moving zero-modes (for instance by assigning consistently a
slightly non-zero conformal dimension to one of the zero-modes and
taking into account only the modes with positive conformal
dimension)\footnote{This is as in the compact case
  \cite{Witten:1993jg}.}.  Thus, in the diagonal subsector, we
have one fermion of charge $-1$, and one bosonic zero-mode of charge
$+1/k$.  The diagonal partition function weighted on the right with
the fermion number and on the left with the $U(1)_R$ charge coded
by the power of $z=e^{2 \pi i \alpha}$ is\footnote{We use the conventions
of \cite{Polchinski} for the $\eta$- and $\theta$-functions.}:
\begin{eqnarray}
\chi_{free,off-diag}&=& \frac{i \theta_{11} (q,z)}{\eta^3} \frac{1}{1-z^{1/k}}.
\end{eqnarray}
Note that we evaluated the elliptic genus without regard to the
presence of the superpotential term.  In a second step, we
re-introduce the winding sectors of the model.

\subsection*{Go with the flow}
We introduce the winding sectors for the compact boson as follows. The
integer left- and right-moving momenta differ by a multiple of
$2k$. The winding sectors can be taken into account by summing
independently over an extra integer $m$ on the left that subtracts $2km$
units of left-moving momentum. We now want to implement that extra sum
in the partition function. To understand an easy way to implement the sum,
it is convenient to temporarily consider the expressions for the
asymptotic $N=2$ superconformal algebra (at large values of $\rho$, where the
potential is negligible).  The asymptotic $N=2$
superconformal algebra is (see e.g. \cite{Murthy:2003es}):
\begin{eqnarray}
T_{as} &=& - \frac{1}{2} (\partial \rho)^2 
- \frac{1}{2} (\partial \theta)^2
- \frac{1}{2} (\psi_\rho \partial \psi_\rho+\psi_\theta \partial \psi_\theta)
- \frac{1}{2} Q \partial^2 \rho \nonumber \\
G^{\pm}_{as} &=& \frac{i}{\sqrt{2}} (\psi_\rho \pm i \psi_\theta)
\partial (\rho \mp i \theta) 
+ \frac{i}{\sqrt{2}} Q \partial(\psi_\rho \pm i \psi_\theta)
\nonumber \\
J^R_{as} &=& i Q \partial \theta - i \psi_\rho \psi_\theta,
\end{eqnarray}
with slope $Q=\sqrt{2/k}$ and central charge $c=3+6/k$. The real boson
$\rho$ parameterizes the asymptotic linear dilaton direction, and the
real field $\theta$ the angular direction at infinity.  We also define
the bosonized complexified fermions $ e^{\pm i H}= \frac{1}{\sqrt{2}} (
\psi_\rho \pm i \psi_\theta )$ and find the algebra in
terms of these variables:
\begin{eqnarray}
T_{as} &=& - \frac{1}{2} (\partial \rho)^2 
- \frac{1}{2} (\partial \theta)^2
- \frac{1}{2} (\partial H)^2
- \frac{1}{2} Q \partial^2 \rho \nonumber \\
G^{\pm}_{as} &=& i e^{\pm i H}
\partial (\rho \mp i \theta) 
+ i Q \partial e^{\pm i H}
\nonumber \\
&=& i e^{\pm i H} \partial (\rho \mp i (\theta- Q H))
\nonumber \\
J^R_{as} &=& i Q \partial \theta + i \partial H.
\label{n=2asymptotic}
\end{eqnarray}
We introduce the extra left-moving momentum $-2km$ in the field
$\theta$ by performing a local $U(1)_R$ transformation (since this is
consistent with the $N=2$ superconformal algebra and makes it easy to
write down the partition sum in the new sector). This is equivalent to
the action of spectral flow on the $N=2$ superconformal
algebra. Explicitly, we see that introducing the quantum number $m$
on the left is equivalent to mapping $\theta \rightarrow \theta -i m
\sqrt{2k} z$ and $H \rightarrow H- i m k z$ where $z$ is the
complexified periodic worldsheet coordinate on the torus. The new
asymptotic algebra then reads:
\begin{eqnarray}
T_{as} &\rightarrow & T_{as} + \frac{c}{6} (km)^2  + km (i Q \partial \theta 
+ i\partial H ) 
\nonumber \\
G^\pm_{as} & \rightarrow & e^{\mp km z} G^{\pm}_{as} 
\nonumber \\
J_{as} & \rightarrow & J_{as} + \frac{c}{3} km.
\end{eqnarray}
The latter transformation is spectral flow by $km$ units. 
Therefore, we can restore the off-diagonal sectors by performing the spectral flow operation on
the holomorphic partition sum (while keeping track of the fact that also the fermion
number shifted by $km$) and we find:
\begin{eqnarray}
\chi_{hol}&=&    \sum_{m \in \mathbb{Z}}
\frac{i \theta_{11} (q,z)}{\eta^3} z^{2m} q^{km^2} 
 \frac{1}{1-z^{1/k} q^m}.
\label{free}
\end{eqnarray}
The interpretation is clear. The right-moving momentum of the angular
mode is still compensated by a diagonal Liouville momentum, but in the
process every unit left-moving momentum mode picks up a conformal
dimension $m$ in sector $m$. The vacuum has
obtained additional left-moving R-charge and it acquired an extra
contribution to its conformal dimension from the extra left-moving
momenta.

\subsection*{A remark on the individual contributions}

It is natural to rewrite the holomorphic partition sum in terms of a higher
  level Appell function $K_l$:
\begin{eqnarray}
K_l (q,z,y) &=& \sum_{m \in Z} \frac{q^{l m^2/2} z^{ml}}{1-yz q^m}.
\end{eqnarray}
We have:
\begin{eqnarray}
\chi_{hol}
&=& \frac{i \theta_{11}(q,z)}{\eta^3} 
K_{2k}(q,z^{1/k},1).
\end{eqnarray}
Above we introduced the winding sectors via a spectral flow operation.
It is interesting to ask how the regularization of the bosonic zero-mode
carries over to the non-diagonal sectors through the spectral flow operation.
Given the choice $|q|<|z^{1/k}|<1$, we can expand the elliptic genus,
and identify the individual state contributions that we have allowed:
\begin{eqnarray}
\chi_{hol}&=& \frac{i \theta_{11}(q,z)}{\eta^3} 
\sum_{m \in \mathbb{Z}} \frac{z^{2m} q^{km^2}}{1-z^{1/k} q^m}
\nonumber \\
&=&
\frac{i \theta_{11}(q,z)}{\eta^3} 
(\sum_{m \ge 0,p \ge 0} - \sum_{m \le -1 ,p \le -1})
q^{km^2+pm} z^{2m} z^{ p/k}
\nonumber \\
 &=&
\frac{i \theta_{11}(q,z)}{\eta^3} 
\big( \sum_{w \le 0, n+kw \ge 0} - \sum_{w \ge 1, n+kw \le -1} \big) q^{-nw} z^{(n-kw)/k}.
\end{eqnarray}
For zero winding, we recuperate the fact that we allowed positive
momentum only. The generalization is that for negative winding, we
allow positive right-moving (angular) momenta, and for strictly
positive winding, we allow strictly negative right-moving momenta. It
is possible to understand this as follows.  We fix the difference of left-
and right-moving momenta in each sector labelled by $w$. We then allow
for diagonal operators only to act, and this includes a dimension
zero operator with tuned Liouville momentum adapted to the
diagonal angular momentum. We pick a right-moving ground state 
and can act with the diagonal operator to
create new right-moving ground states. It is the conformal dimension
of the diagonal operators that is regularized in each sector, and
therefore, the sign of the right-moving momentum that determines our
regularization scheme. The left-moving momentum, on the other hand,
can be non-zero since the left-movers are not necessarily in the
ground state. It is set by the diagonal operator, and the sector
label $w$, for a total momentum of $n-kw$ on the left.

\subsection{The algebraic perspective}
We want to compare the final expression we obtained in the free field
analysis with a proposal \cite{Eguchi:2004yi} that has an algebraic
origin \cite{Hanany:2002ev,Eguchi:2004yi}.  The higher level Appell
function is a sum over extended twisted Ramond sector characters
\cite{Eguchi:2004yi} for an $N=2$ superconformal theory with central
charge of the form $c=3+\frac{6}{k}$ with $k$ a positive
integer. Recall that the extended twisted R-sector characters are \cite{Eguchi:1987sm}:
\begin{eqnarray}
Ch_d^{\tilde{R}} (j,r';q,z) &=&
\sum_m \frac{i \theta_{11} (q,z)}{\eta^3}
\frac{1}{1-z q^{k(m+ (2r'+1)/2k)}}
\nonumber \\
& &
z^{\frac{2j-1}{k}} q^{k (m + \frac{2r'+1}{2k})(2j-1)/k}
z^{2(m + \frac{2r'+1}{2k})} q^{k (m + \frac{2r'+1}{2k})^2}.
\end{eqnarray}
When we evaluate the character for a representation built on a ground
state with $r'=-1/2$ (which gives rise to zero conformal dimension for
the value $m=0$ in the infinite sum), and we sum over spins $j$ that
lie in the set $\{ 1/2, 1, \dots, k/2 \}$ and which appear in a
decomposition of the partition function
\cite{Hanany:2002ev,Eguchi:2004yi,Israel:2004ir}, we find
the following result \cite{Eguchi:2004yi}:
\begin{eqnarray}
\sum_{2j-1=0}^{k-1} Ch_d^{\tilde{R}} (j,-1/2;q,z)
&=&\frac{i \theta_{11} (q,z)}{\eta^3} 
 \sum_m z^{2m } q^{k m^2}
\frac{1}{1-z q^{k m}} \sum_{2j-1=0}^{k-1}
z^{\frac{2j-1}{k}} q^{m(2j-1)}
\nonumber \\
&=&\frac{i \theta_{11} (q,z)}{\eta^3}  \sum_m z^{2m} q^{k m^2}
\frac{1}{1-z q^{k m}} \frac{1-z q^{km}}{1-z^{1/k} q^m} 
\nonumber \\
\nonumber \\
&=& \chi_{hol}.
\label{algebraic}
\end{eqnarray}
It is gratifying that the simple free field derivation agrees with the
algebraic perspective.  The algebraic result is based on the idea that
only a definite range of spins contribute to the elliptic genus, and that
the elliptic genus is made of extended characters, i.e. that all
spectrally flowed Hilbert spaces must be taken into account. This is consistent with the
analysis of the partition function
\cite{Hanany:2002ev,Eguchi:2004yi,Israel:2004ir}. Moreover, there are
various string theory inspired consistency checks on these assumptions
ranging from consistency with linear dilaton holography
\cite{Giveon:1999px} to the physics of strings puffing up in $AdS_3$
\cite{Maldacena:2000hw}.

\subsection*{Quandary}
The free field and the algebraic perspective on the holomorphic part
of the elliptic genus are well-motivated and give the same result, yet
they cannot tell us the whole story. The modular transformation
properties of the higher level Appell function supplemented with the
$\theta$- and $\eta$-functions (see e.g. \cite{Semikhatov:2003uc}) do
not agree with the expectation that the elliptic genus be a Jacobi
form. The anomaly is associated to the non-compactness of target space
which contaminates the above derivation with a volume divergence
(that is absent in the compact case). In the following, we will
carefully regularize the volume divergence to gain further insight
into the modular, or holomorphic anomaly.

\section{The path integral and modularity}
Since the path integral (or Lagrangian) formulation of the elliptic
genus manifestly codes its modular transformation properties, it will
be interesting to compare the path integral calculation to the
holomorphic perspectives given in the previous section. We will
perform the full path integral  (in contrast to the compact case,
where it was computed through localization, which reduces the
calculation to the free field calculation
\cite{Henningson:1993nr}). Most of the details of the derivation of
the path integral expression are identical to those provided in
\cite{Hanany:2002ev}. Furthermore, we draw upon the analysis in the compact case
\cite{Henningson:1993nr} to properly implement the twisting by the
left $U(1)_R$ charge.

A careful discussion of the calculation of the elliptic genus in the
Lagrangian formulation of compact $N=2$ superconformal
Wess-Zumino-Witten models was given in \cite{Henningson:1993nr}. The
analysis is mostly valid for non-compact target space groups as
well. If we introduce an $SL(2,\mathbb{R})$ group valued field $g$,
two (right- and left-moving) fermions $\chi^{\pm}$ which take values
in the algebra $sl(2,\mathbb{R})$ (mod $u(1)$), and a generator $U$ of the $U(1)$
gauge group, then the fields are argued to transform under the
left-moving $U(1)_R$ symmetry with parameter $\gamma$ as:
\begin{eqnarray}
\delta \chi_+ &=& -\frac{i \gamma}{k} \chi_+
\nonumber \\
\delta \chi_- &=& \frac{i \gamma (k+1) }{k} \chi_-
\nonumber \\
\delta g &=& \frac{i \gamma}{ k} (Ug-gU).
\end{eqnarray}
We will not review the derivation of these formulas, based on the 
chiral anomalies of the two-dimensional model, but we indicate
the main differences with the compact case. When we compare to
equation $(25)$ of \cite{Henningson:1993nr}, we have made the
following changes. By convention, we work with the supersymmetric
level $k$ of the model, namely the level that takes into account both
bosons and fermions\footnote{Therefore for a compact model our level
  $k^{SU(2)}$ would be related to the level $k_{Hen}$ as $k^{SU(2)}=
  k_{Hen}+2$.}.  Moreover we work with a non-compact model which
formally corresponds to changing the sign of the total level
$k$.
When we take those changes into account,
we find the formulas quoted above from the analysis of
\cite{Henningson:1993nr} adapted to an axially gauged non-compact model. We conclude
that in some suitable normalizations the charges of the 
right-moving fermion
$\chi_+$ and of the (non-Cartan) complex boson agree (and are $1/k$), while the
charge of the other fermion is bigger by a factor $k+1$. 

The other ingredient we use is the path integral with ordinary
boundary conditions in the NSNS sector of the theory as computed in
\cite{Hanany:2002ev,Eguchi:2004yi,Israel:2004ir}. It consists of a
contribution from the non-compact coset, from the fermions, and from
Wilson lines $s_{1,2}$ for the gauge field on the torus. However, in
comparison to \cite{Hanany:2002ev,Eguchi:2004yi,Israel:2004ir}, we
make the following changes. It will be more intuitive to gauge the
direction T-dual to the one gauged in those papers. We must also take into account the fact that the
right-moving boson and fermion oscillators cancel out against
each other. The result of the path integral calculation of the elliptic
genus (which in the coordinate choice found in \cite{Gawedzki:1991yu}
reduces to the evaluation of a Ray-Singer torsion, a twisted fermion partition function,
and zero modes) is then:
\begin{eqnarray}
\chi &=& \sum_{m,w} \int_{0}^{1} ds_1 ds_2 
\frac{\theta_{11} (\tau,  s_1 \tau + s_2- \frac{k+1}{k}\alpha )}{ \theta_{11} (\tau, s_1 \tau + s_2-\frac{1}{k} \alpha)} e^{2 \pi i \alpha w/k}
e^{ - \frac{\pi}{k \tau_2} | (m+ks_2)+\tau (w+k s_1)|^2}.
\label{first}
\end{eqnarray}
The $\theta_{11}$ functions have a twisted argument that depends on
the holonomies of the gauge field on the torus, and on the twist by
the $U(1)_R$ charges, with a weight determined by the charges of the
left-moving bosons and fermions (analyzed above). The last factor codes the
coupling of the oscillators to the bosonic zero-modes at radius
$R=\sqrt{2/k}$ via the holonomies $s_{1,2}$. The twist of the
bosonic zero-modes also introduces and extra phase factor $e^{ 2 \pi i \alpha w/k}$.

Note that the result is not holomorphic. This possibility is opened up
by the cancellation of bosonic and fermionic oscillators on the right,
in the presence of further zero modes. The cancellation is between a
regulated right-moving fermionic zero-mode and the volume divergence.
Moreover, the expression remains formal at this stage, since at non-zero
twist $\alpha$, there are poles in the theta-function in the
denominator that are not compensated by zeroes in the numerator (in
contrast to the case $\alpha=0$, where the calculation gives a Witten
index equal to $k$). Thus, the integral and the result are still
divergent. We will provide a further regularization.

\subsection*{A first look at modularity}
In a first step though we check the modular transformation
properties of the formal integral expression for the elliptic
genus. We show that the path integral result
satisfies the expected modular covariance properties. We have 
a Jacobi form of weight $0$ and index $k(k+2)/2=k^2 c/6$ \footnote{
Strictly speaking, for odd level $k$, the index is $2k(k+2)$.}.  

The expected modular covariance properties are \cite{Kawai:1993jk},
from the boundary conditions on the path integral and the factorization
of the $U(1)_R$ current algebra:
\begin{eqnarray}
\chi(\tau+1, \alpha) &=& \chi(\tau,\alpha)
\nonumber \\
\chi(-\frac{1}{\tau},\frac{\alpha}{\tau}) &=& e^{2 \pi i \frac{c}{6} \alpha^2/\tau}
\chi(\tau,\alpha).
\end{eqnarray}
If all $U(1)_R$ charges in the NS sector are multiples of $1/k$,
then we expect, with $\mu \in k \mathbb{Z}$:
\begin{eqnarray}
  \chi(\tau,\alpha+\mu) &=& (-1)^{\frac{c}{3} \mu} \chi(\tau,\alpha)
\end{eqnarray}
and from mapping the Ramond sector states into the Ramond sector states
after an integer number of spectral flows (with $\lambda \in k \mathbb{Z}$):
\begin{eqnarray}
\chi(\tau, \alpha + \lambda \tau) &=& (-1)^{\frac{c}{3} \lambda} 
e^{- 2 \pi i \frac{c}{6} (\lambda^2 \tau + 2 \lambda \alpha)} \chi(\tau,\alpha).
\end{eqnarray}
Finally, we have
\begin{eqnarray}
\chi(\tau,\alpha) &=& \chi(\tau,-\alpha)
\end{eqnarray}
for a charge conjugation invariant Ramond spectrum.
To check the modular properties, it is convenient to first doubly Poisson resum the path integral
result. We obtain:
\begin{eqnarray}
\chi &=& k \sum_{m,w} \int_{0}^{1} ds_1 ds_2 
\frac{\theta_{11} (\tau,  s_1 \tau + s_2- \frac{k+1}{k}\alpha )}{ 
\theta_{11} (\tau, s_1 \tau + s_2-\frac{1}{k} \alpha)} 
e^{-2 \pi i s_2 kw } e^{2 \pi i s_1 (km-\alpha) } 
e^{ - \frac{\pi k}{ \tau_2} | (m-\frac{\alpha}{k})+\tau w|^2}.
\label{second}
\end{eqnarray}
Under the simultaneous transformations (implied
by following the holonomies and winding numbers under the modular
action):
\begin{eqnarray}
& & \tau  \rightarrow  \tau+1 
\qquad \quad 
s_2  \rightarrow s_2 - s_1
\qquad \quad
m  \rightarrow  m-w,
\end{eqnarray}
the elliptic genus is invariant. Under the transformations:
\begin{eqnarray}
& & 
\tau  \rightarrow  - \frac{1}{\tau}
\qquad
s_1  \rightarrow  -s_2
\qquad
w  \rightarrow  -m
\nonumber \\
& & 
\alpha \rightarrow  \frac{\alpha}{\tau}
\qquad \quad
s_2  \rightarrow  s_1 
\qquad
\quad
m  \rightarrow  w,
\end{eqnarray}
we pick up the expected factor:
\begin{eqnarray}
e^{ \pi i ( (s_1 \tau+s_2 - \frac{k+1}{k} \alpha)^2
- (s_1 \tau+s_2 - \frac{\alpha}{k})^2)/\tau} 
e^{2 \pi i \alpha s_2/\tau} e^{2 \pi i s_1 \alpha}
&=& e^{ \pi i/\tau \alpha^2 (1 + \frac{2}{k})}. 
\end{eqnarray}
One similarly verifies that the elliptic
genus satisfies the periodicity and parity requirements as well.
These properties make the elliptic genus $\chi$ a
Jacobi form of weight zero and index $k^2 c/6=(k+2)k/2$,
at least formally. 

\subsection*{Distinguishing two contributions}
The path integral result can be related to the holomorphic perspective
through an analysis similar to
\cite{Hanany:2002ev,Eguchi:2004yi,Israel:2004ir} while keeping track
of the degeneracies of descendent states as in \cite{Israel:2004ir}. We perform the
following manipulations on the path integral result.  We singly Poisson
resum on $m$ in equation (\ref{second}) to find:
\begin{eqnarray}
\chi 
&=&  \sqrt{k \tau_2} \sum_{n,w} \int_0^1 d s_1 
\int_0^1 d s_2 \frac{\theta_{11} (\tau,  s_1 \tau + s_2- \frac{k+1}{k}\alpha )}{ \theta_{11} (\tau, s_1 \tau + s_2-\frac{1}{k} \alpha)} e^{2 \pi i \alpha \frac{ n}{k}
- 2 \pi i s_2 kw}
q^{(kw-(n+k s_1))^2/4k} \bar{q}^{(kw+(n+k s_1))^2/4k}.
\nonumber
\end{eqnarray}
We now regularize the path integral by  assuming that
$1> |q^{s_1} z^{-\frac{1}{k}} > |q|$. We take $|z|$ very close to one,
and  cut off the boundaries
of the integration
over the holonomy $s_1$ such as to satisfy the above equation. 
This regularization allows us to expand the
$\theta_{11}$ function in the denominator in terms of the
special functions $S_r(\tau)$ which are known to code the degeneracies of
descendant states in the discrete characters of the $SL(2,R)/U(1)$ coset
conformal field theory
\cite{Sfetsos:1991wn,Bakas:1991fs,Griffin:1990fg,Pakman:2003kh}.
The definition of the series $S_r$ (which are related
to Hecke indefinite modular forms) is:
\begin{eqnarray}
S_r (\tau) &=& \sum_{n=0}^{+\infty} (-1)^n q^{ \frac{n(n+2r+1)}{2}}.
\end{eqnarray}
After the expansion, we obtain the expression:
\begin{eqnarray}
\chi &=&  -\sqrt{k \tau_2} \frac{1}{\eta^3}
\sum_{m,n,w,r} \int ds_{1} \int ds_{2}
(-1)^m q^{(m-\frac{1}{2})^2/2} (z^{1 + \frac{1}{k}} e^{-2 \pi i s_2}
q^{-s_1})^{m-1/2}
(e^{ 2 \pi i s_2} q^{s_1} z^{-\frac{1}{k}})^{r + \frac{1}{2}} S_r
\nonumber \\
& & e^{- 2 \pi i s_2 kw} z^{\frac{n}{k}}
q^{(kw-(n+k s_1))^2/4k} \bar{q}^{(kw+(n+k s_1))^2/4k}.
\end{eqnarray}
The integral over $s_2$ implies that we must have 
$m-r -1+ kw= 0$ to get a non-zero contribution.
We shift the summation variable
$n$ to $v=n+kw$ and find:
\begin{eqnarray}
\chi &=&  -\sqrt{k \tau_2} \frac{1}{\eta^3}
\sum_{m,v,w,r} \int ds_{1} \int ds_{2}
(-1)^m q^{(m-\frac{1}{2})^2/2} 
z^{m-\frac{1}{2}} S_r 
\nonumber \\
& & 
z^{-2w+\frac{v}{k}} e^{- 2 \pi i s_2 (m-r-1+kw)}
q^{kw^2-vw}
(q \bar{q})^{(v+ks_1)^2/4k}.
\end{eqnarray}
We can then introduce 
an integral over continuous momenta $s$ to make the
exponent linear in $s_1$:
\begin{eqnarray}
\chi &=& - 2 \tau_2 \frac{1}{\eta^3}
\sum_{m,v,w,r} \int ds_{1} \int ds_2 \int_{- \infty}^{+\infty} ds
(-1)^m q^{(m-\frac{1}{2})^2/2} z^{m-\frac{1}{2}}
S_r z^{-2w+\frac{v}{k}} q^{kw^2-vw} \nonumber \\
& & e^{ 2 \pi i s_2 (r-kw-m+1)}
(q \bar{q})^{s_1(is+\frac{v}{2})+\frac{s^2}{k}+\frac{v^2}{4k}},
\end{eqnarray}
which allows us to easily perform the integrations over both holonomies
$s_{1,2}$:
\begin{eqnarray}
\chi &=& \frac{1}{\pi} \frac{1}{\eta^3} \sum_{m,v,w} 
\int_{\mathbb{R}-i \epsilon} \frac{ds}{2 i s + v}
(-1)^m q^{(m-\frac{1}{2})^2/2} z^{m-\frac{1}{2}}
(q \bar{q})^{\frac{s^2}{k}+\frac{v^2}{4k}}
 z^{-2w+\frac{v}{k}} q^{kw^2-vw} 
\nonumber \\
& &
S_{kw+m-1}
( (q \bar{q})^{is + \frac{v}{2}}-1). 
\end{eqnarray}
Note that we slightly shifted the integration over the momentum $s$ off the real axis for future
convenience.
Drawing inspiration from \cite{Hanany:2002ev} on how to disentangle contributions
with continuous radial momentum from discrete contributions,
we now distinguish between two parts in the path integral.
One which we will call the
remainder term $\chi_{rem}$, and a second one 
which will turn out to
be holomorphic:
\begin{eqnarray}
\chi &=& \chi_{hol} + \chi_{rem}
\nonumber \\
\chi_{hol} &=&  \frac{1}{\pi} \frac{1}{\eta^3} \sum_{m,v,w} 
\int_{\mathbb{R}-i \epsilon} \frac{ds}{2 i s + v}
(-1)^m q^{(m-\frac{1}{2})^2/2} z^{m-\frac{1}{2}}
(q \bar{q})^{\frac{s^2}{k}+\frac{v^2}{4k}}
 z^{-2w+\frac{v}{k}} q^{kw^2-vw} 
\nonumber \\
& &
(1-S_{kw+m-1}
(1- (q \bar{q})^{is + \frac{v}{2}})
\nonumber \\
\chi_{rem} &=& -\frac{1}{\pi} \frac{1}{\eta^3} \sum_{m,v,w} 
\int_{\mathbb{R}-i \epsilon} \frac{ds}{2 i s + v}
(-1)^m q^{(m-\frac{1}{2})^2/2} z^{m-\frac{1}{2}}
(q \bar{q})^{\frac{s^2}{k}+\frac{v^2}{4k}}
 z^{-2w+\frac{v}{k}} q^{kw^2-vw}.
\label{remetal}
\end{eqnarray}
We first massage the holomorphic part, using the property
 $S_r=S_{-r-1}-1$ and shifting the summation variable $m$
by one:
\begin{eqnarray}
\chi_{hol} &=&  \frac{1}{\pi} \frac{1}{\eta^3} \sum_{m,v,w} 
\int_{\mathbb{R}-i \epsilon} \frac{ds}{2 i s + v}
(-1)^m q^{(m-\frac{1}{2})^2/2} z^{m-\frac{1}{2}}
(q \bar{q})^{\frac{s^2}{k}+\frac{v^2}{4k}}
 z^{-2w+\frac{v}{k}} q^{kw^2-vw}
\nonumber \\
& & 
(S_{-m-kw}-z q^{m+is +\frac{v}{2}} \bar{q}^{is +\frac{v}{2}} S_{kw+m}) 
\end{eqnarray}
and using furthermore the formula $q^r S_r=S_{-r}$ we find:
\begin{eqnarray}
\chi_{hol} &=&  \frac{1}{\pi} \frac{1}{\eta^3} \sum_{m,v,w} 
\int_{\mathbb{R}-i \epsilon} \frac{ds}{2 i s + v}
(-1)^m q^{(m-\frac{1}{2})^2/2} z^{m-\frac{1}{2}}
(q \bar{q})^{\frac{s^2}{k}+\frac{v^2}{4k}}
 z^{-2w+\frac{v}{k}} q^{kw^2-vw} S_{-m-kw}
\nonumber \\
& & 
(1-z q^{-kw+is +\frac{v}{2}} \bar{q}^{is +\frac{v}{2}}).
\end{eqnarray} 
The neat property of the second term in this expression is that it can
be rewritten as an integral over the real axis, shifted upward by $k/2$,
if we simultaneously shift the momentum $v$:
\begin{eqnarray}
\chi_{hol} &=&  \frac{1}{\pi} \frac{1}{\eta^3} \sum_{m,v,w} 
(\int_{\mathbb{R}-i \epsilon} - \int_{\mathbb{R}+ i \frac{k}{2}-i \epsilon}
) \frac{ds}{2 i s + v}
(-1)^m q^{(m-\frac{1}{2})^2/2} z^{m-\frac{1}{2}}
(q \bar{q})^{\frac{s^2}{k}+\frac{v^2}{4k}}
 z^{-2w+\frac{v}{k}} q^{kw^2-vw} S_{-m-kw}. \nonumber 
\end{eqnarray} 
Splitting the integral, and performing the spectral flow operation has
disentangled discrete characters from the continuum.
Indeed, the good convergence properties at infinity allow us to
interpret our formula for $\chi_{hol}$ as a contour integral that picks up the
poles in $2is+v$ that lie between $-i \epsilon$ and $ik/2-i
\epsilon$. The poles lie at Liouville momenta that give rise to right-moving
ground states. After some further algebraic operations, this gives the 
announced holomorphic result:
\begin{eqnarray}
\chi_{hol} &=& \frac{1}{\eta^3} \sum_{2j-1=0}^{k-1} \sum_{w,m}
 (-1)^m q^{(m-\frac{1}{2})^2/2} z^{m-\frac{1}{2}} 
z^{-2w}q^{kw^2} S_{-m-kw} (z^{-\frac{1}{k}} q^{w})^{2j-1}
\nonumber \\
&=& \frac{1}{\eta^3} \sum_{m \in \mathbb{Z}} \frac{i \theta_{11}(q,z)}{1-z^{\frac{1}{k}} q^m} z^{2m} q^{k m^2}.
\end{eqnarray}
Thus we see that the path integral does partially agree with the
intuitive derivation of the (holomorphic part of the) elliptic genus from
the free field (see equation (\ref{free})) and algebraic perspectives (see equation (\ref{algebraic})).
\subsection*{The remainder term: from mock theta-functions to Jacobi forms} 
We do also have the remainder term $\chi_{rem}$. The expression
(\ref{remetal}) for the remainder term can be read as a trace over the
asymptotic continuum of states with weighting by the left
$U(1)_R$-charge and the conformal weights, in which the right-moving
oscillators of fermions and bosons have cancelled out.
It therefore also 
provides a Hamiltonian view of
the holomorphic anomaly.  The
cancellation has left a measure factor which is given by one over the
total right-moving momentum, which is formally the ratio of the
right-moving fermionic zero mode and a right-moving supercharge (as
can be seen from the analogue of equation (\ref{n=2asymptotic}) for
the right-moving superconformal algebra).  The measure is such that
the contributions which localize in momentum space are
holomorphic. However, the integral over the continuum associated to
the non-compactness of the target space is non-zero and
non-holomorphic. 

We evaluate the integral over the continuum
of momenta $s$ to obtain:
\begin{eqnarray}
\chi_{rem} &=& - \frac{i \theta_{11}(q,z)}{2 \eta^3}
\sum_{v,w} z^{-2 w+ \frac{v}{k}} q^{k w^2 - vw}
( \mbox{sgn} (v+\epsilon) - \mbox{Erf}( v \sqrt{\frac{\pi \tau_2}{k}})),
\end{eqnarray}
where $\mbox{Erf}$ is the error function. The role of the remainder term in the path integral
is to render the elliptic genus modular covariant. Indeed,
we can now make our analysis of the modular transformation properties
rigorous by comparing our results to techniques developed in the mathematics
literature in the context of the definition and analysis of mock theta
functions. See e.g. \cite{Zagier} for a review and
\cite{Eguchi:2010xk} for a  recent application to the entropy of
Calabi-Yau manifolds. Here we closely follow the presentation of \cite{Zwegers}
where the definition of the level $l$ Appell-Lerch sums is taken to be:
\begin{eqnarray}
A_l(u,v;\tau) &=& a^{l/2} \sum_{n \in \mathbb{Z}} 
\frac{(-1)^{ln}
q^{ln(n+1)/2} b^n}{1-a q^n}
\end{eqnarray}
where $a=e^{2 \pi i u}$ and $b=e^{2 \pi i v}$. It is known that there exists
a
correction term for the Appell function that completes it into a 
Jacobi form  of weight $1$ and index $\left( \begin{array}{cc} -l & 1 \\
1 & 0 \end{array} \right)$. The Jacobi form is the Appell sum plus a remainder
term:
\begin{eqnarray}
\hat{A}_l(u,v;\tau) &=&A_l(u,v;\tau)
+ \frac{i}{2l} a^{(l-1)/2} \sum_{m \, mod \, l} \theta_{11} (\frac{v+m}{l}+\frac{l-1}{2l} \tau; \frac{\tau}{l}) R(u - \frac{v+m}{l}-\frac{l-1}{2l} \tau;
\frac{\tau}{l})
\end{eqnarray} 
where
\begin{eqnarray}
R(u;\tau) &=& \sum_{\nu \in \mathbb{Z}+ \frac{1}{2}}
( \mbox{sgn} (\nu) - \mbox{Erf}(\sqrt{2 \pi \tau_2}(\nu + \mbox{Im} u/ \tau_2))
(-1)^{\nu - \frac{1}{2}} a^{- \nu} q^{- \frac{\nu^2}{2}}.
\end{eqnarray}
We first express the holomorphic part of the elliptic genus as proportional to 
an Appell-Lerch sum:
\begin{eqnarray}
\chi_{hol} &=& z^{-1} \frac{i \theta_{11}(q,z)}{\eta^3} A_{2k}(z^{\frac{1}{k}},z^2 q^{-k};q).
\end{eqnarray}
It is now a matter of straightforward calculation to show that our remainder term
$\chi_{rem}$ precisely agrees with the correction term of \cite{Zwegers} when we
evaluate the latter at $l=2k, \, a=z^{1/k}, \, b= q^{-k} z^2$ and
multiply by $z^{-1} i \theta_{11}/\eta^3$. The fact that $\hat{A}_{2k}$ is a Jacobi
form 
was rigorously proven. We have the modular
properties \cite{Zwegers}:
\begin{eqnarray} 
\hat{A}_{l} (u+1,v) &=& (-1)^l \hat{A}_l(u,v)
\qquad \qquad \qquad \, \, \, 
 \hat{A}_{l} (u,v+1) =  \hat{A}_l(u,v)
\nonumber \\
\hat{A}_{l} (u+\tau,v) &=& (-1)^l a^l b^{-1} q^{\frac{l}{2}} \hat{A}_l(u,v)
\qquad \quad
\hat{A}_{l} (u,v+\tau) =  a^{-1} \hat{A}_l(u,v)
\nonumber \\
\hat{A}_{l}(u,v;\tau+1) &=& \hat{A}_l (u,v;\tau)
\qquad \qquad \qquad \quad \,
\hat{A}_l (\frac{u}{\tau},\frac{v}{\tau};-\frac{1}{\tau})
= \tau e^{\pi i (2v-lu)u/\tau} \hat{A}_l (u,v;\tau). 
\end{eqnarray}
Thus the chosen regularization of the path integral is indeed modular
covariant. The non-holomorphic elliptic genus as coded in the
regularized path integral is a Jacobi form with weight zero and index
$k(k+2)/2$.  Multiplication by the $\theta$- and $\eta$-function
indeed gives the expected weight and index to the elliptic genus. To
check this it is useful to use the periodicity of the generalized
Appell-Lerch sum and write:
\begin{eqnarray}
\chi &=& \frac{i \theta_{11}(q,z)}{\eta^3} \hat{A}_{2k} (z^{\frac{1}{k}},z^2; q),
\end{eqnarray}
and then use the modular covariance properties listed above to complete the proof
of modularity.

\section{Conclusions}
In this paper we gave a free field derivation of the holomorphic part
of the elliptic genus of a basic $N=2$ superconformal field theory
with central charge $c$ larger than three and equal to
$c=3+\frac{6}{k}$ with $k$ integer. The holomorphic part of the
elliptic genus is directly related to an Appell-Lerch sum. We also
explicitly evaluated the path integral over the coset conformal field
theory that corresponds to the non-compact Landau-Ginzburg model. We
showed that it contains a holomorphic part which agrees with the free
field and the algebraic analysis.  The full path integral result though is 
non-holomorphic and modular. The holomorphic anomaly satisfied by the elliptic
genus finds its origin in the non-compactness of target space. The
path integral provides a physical origin for the
remainder term postulated in the mathematics literature to complete 
mock theta-functions into  Jacobi forms.  Our result 
provides a modular covariant regularization to a problem plagued by a
volume divergence that cannot be factored out as the volume of a
symmetry group. It gives rise to an expression that can consistently
be integrated over moduli space (e.g. in string amplitudes).

There are many directions that open up thanks to this elementary
result. We mention only three. The derivation can be extended to $N=4$
models, to combinations of compact and non-compact $N=2$ or $N=4$
models, and orbifolds thereof. Secondly, we can analyze further the
holomorphic anomaly differential equation that is satisfied by the
remainder term (when we derive with respect to $\bar{\tau}$).
Thirdly, we can apply the insight we gained into modular covariant
regularization to the bulk modular invariant partition function (of
the bosonic or supersymmetric coset model).

\section*{Acknowledgements}
We would like to thank Sujay Ashok, Costas Bachas, Raphael Benichou,
Tohru Eguchi, Dan Israel, Ari Pakman and Giuseppe Policastro for
useful discussions.  Our work was supported in part by the grant
ANR-09-BLAN-0157-02.

\end{document}